\documentstyle[12pt]{article}
\begin{document}

\vspace*{2cm}

\noindent{\large\bf Colloids with polymer stars: The interaction}

\vspace{1.5ex}

\noindent
Ch. von Ferber$^a$, Yu. Holovatch$^b$, A.~Jusufi$^a$ C.N.~Likos$^a$,
H.~L\"owen$^a$ and M.~Watzlawek$^a$
\vspace{1.5ex}

\noindent
$^a$
Institut f\"ur Theoretische Physik II, Heinrich-Heine-Universit\"at
D\"usseldorf, \\
D-40225 D\"usseldorf, Germany

\vspace{1.5ex}

\noindent
$^b$
Institute for Condensed Matter Physics, Ukrainian Acad. Sci.,\\
1 Svientsitskii Str., UA-290011, Lviv, Ukraine

\vspace{2ex}

We derive the short distance interaction of star polymers in
a colloidal solution.
We calculate the corresponding force between two stars with arbitrary
numbers of legs $f_1$ and $f_2$.
We show that a simple scaling theory originally derived
for high $f_1,f_2$ nicely
matches with the results of elaborated renormalization
group analysis for $f_1+f_2\leq 6$ generalizing and
confirming a previous conjecture based only on scaling
results for $f_1=f_2=1,2$.

\vspace{2ex}

Star polymers, i.e. structures of polymer chains that are chemically linked
with one end to a common core, have found recent interest as very soft
colloidal particles \cite{Likos,Semenov98}.
Increasing the number of chains $f$ they interpolate between the properties
of linear polymers and polymeric micelles \cite{Gast}.
For large $f$ the effective interaction between the cores of different
polymer stars becomes strong enough to allow for crystal ordering of a
dense solution of stars. While
such a behavior was already predicted by early scaling arguments
\cite{WittenPincus} only recently corresponding experiments have become
feasible with sufficiently dense stars. In addition theory and computer
simulation have refined the original estimate for the
number of chains $f$ necessary for a freezing transition from $f\sim 100$ to
$f\sim 34$ and predicted a rich phase diagram \cite{Likos}.
These results were derived using an effective pair potential between stars
\cite{Likos}
with a short distance behavior derived from scaling theory as explained below.

The scaling theory of polymers was significantly advanced by de Gennes
observation that the $n$-component spin model of magnetic systems applies to
polymers in the formal $n=0$ limit \cite{deGennes71}.
This opens the way to apply renormalization group (RG) theory to explain the
scaling
properties of polymer solutions that have been the subject of experimental and
theoretical investigations since the pioneering works in this field
\cite{nobel}.
Many details of the behavior of polymer solutions may be derived using the RG
analysis \cite{Schaefer99}. Here, we use only the more basic results of power
law scaling: The radius of gyration $R_g(N)$ of a polymer chain and the
partition
sum ${\cal Z}(N)$ are found to obey the power laws
\begin{equation}
R_g(N)\sim N^\nu \mbox{\hspace{3em} and \hspace{3em}}
{\cal Z}(N)\sim z^N N^{\gamma-1}
\end{equation}
with some fugacity $z$ that shows the number of possibilities to locally
add one more monomer to the chain, while the two exponents $\nu$ and $\gamma-1$
represent the nontrivial corrections that are due to the self-avoiding
interaction that is non-local along the chain.
These exponents are the $n=0$ limits of the correlation length exponent
 $\nu(n)$ and the susceptibility exponent $\gamma(n)$ of the
$n$-component model.
The exponents of any other power law for linear polymers may
be expressed by these two exponents in terms of scaling relations.

It has been shown that the $n=0$-component spin model can be extended
to describe polymer networks and in particular star polymers
\cite{Schaefer92}. A family of additional exponents $\gamma_f$
governs the scaling of the partition sums ${\cal Z}_f(N)$ of
polymer stars of $f$ chains each with $N$ monomers:
\begin{equation}
{\cal Z}_f(N)\sim z^{fN} N^{\gamma_f-1} .
\end{equation}
Again the exponents of any other power law for more general polymer
networks is given by scaling relations in terms of $\gamma_f$ and $\nu$.
For details of a proof and necessary restrictions on the chain length
distributions see \cite{Schaefer92}.
For large $f$ each chain of the star is restricted approximately to
a cone of space angle $\Omega_f=4\pi/f$. In this cone approximation
one finds for large $f$ \cite{Ohno88}
\begin{equation}
\gamma_f-1\sim -f^{3/2}.
\end{equation}
Let us now turn to the effective interaction between the cores of
two star polymers at small distance (small on the scale of the
size $R_g$ of the star).
In the formalism of the $n=0$ component model, the core of a
star polymer corresponds to a local product of $f$ spin fields
$\phi_1({\bf x})\cdots\phi_f({\bf x})$, each representing the endpoint
of one polymer chain.
The probability of approach of the cores of two star
polymers at small distance $r$ results in these terms from a
short distance expansion for the composite fields.
Let us consider the general case of two stars of different functionalities
$f_1$ and $f_2$.
The power law for the partition sum ${\cal Z}^{(2)}_{f_1f_2}$
of two such star polymers at distance $r$ \cite{Duplantier88}
\begin{equation}
{\cal Z}^{(2)}_{f_1f_2}(r)\sim r^{\Theta^{(2)}_{f_1f_2}}
\end{equation}
is governed by the contact exponent $\Theta^{(2)}_{f_1f_2}$.
Then the short distance expansion provides the scaling relations
\begin{eqnarray}
\nu\Theta_{f_1f_2} &=&
(\gamma_{f_1}-1)+(\gamma_{f_2}-1)-(\gamma_{f_1+f_2}-1)\,,
\nonumber\\
\Theta_{f_1f_2} &=&  \eta_{f_1}+\eta_{f_2}-\eta_{f_1+f_2}\,.
\end{eqnarray}
Here, we have substituted an equivalent family of exponents
$\eta_f$ to replace $\gamma_f-1=\nu(\eta_f-\eta_2)$.
The mean force $F^{(2)}_{f_1f_2}(r)$ between two star polymers at short
distance
$r$ is easily derived from the effective potential
$V^{\rm eff}_{f_1f_2}(r)=\log {\cal Z}^{(2)}_{f_1f_2}(r)$ as
\begin{equation}
 F^{(2)}_{f_1f_2}(r)=
\frac{\Theta^{(2)}_{f_1f_2}}{r}\,,
\mbox{ with }\,\Theta^{(2)}_{ff}  \approx\frac{5}{18}f^{3/2} .
\end{equation}
The factor $5/18$ is found by matching the cone approximation for
$\Theta^{(2)}_{ff}$
to the known values of the contact exponents for $f=1,2$ \cite{Likos}.
This matching in turn proposes an approximate value for the $\eta_f$
exponents,
\begin{equation}
 \eta_f\approx -\frac{5}{18} (2^{3/2}-2)f^{3/2} .
\end{equation}
Note that this is inconsistent with the exact result $\eta_1=0$. However, this
assumption nicely reproduces the contact exponents as derived from
3-loop perturbation theory \cite{Ferhol} as is displayed in table 1.

\begin{table}
\caption {
\label{tab1}
The prefactor $\Theta_{f_1f_2}$ of the force between two star polymers
at short distance calculated in non-resummed 1-loop and resummed
3-loop RG analysis in comparison to the result of the cone approximation.
}

\vspace{1.5ex}

\begin{tabular}{lrrrrrl}
\hline
$\Theta_{f_1f_2}$ & $f_2=1$ & $f_2=2$ & $f_2=3$ & $f_2=4$ & $f_2=5$ & app.\\
\hline
$f_1=1$           & -.21    &   -.42  &   -.63  &  -.85   &  -1.06  & 1-loop\\
$f_1=1$           & -.28    &   -.48  &   -.62  &  -.76   &   -.87  & 3-loop\\
$f_1=1$           & -.27    &   -.45  &   -.60  &  -.73   &   -.84 & cone\\
\\
$f_1=2$           &         &   -.85  &  -1.27  & -1.70   &         & 1-loop\\
$f_1=2$           &         &   -.82  &  -1.10  & -1.35   &         & 3-loop\\
$f_1=2$           &         &   -.78  &  -1.05  & -1.29   &        &cone\\
\\
$f_1=3$           &         &         &  -1.91  &         &         & 1-loop\\
$f_1=3$           &         &         &  -1.49  &         &         & 3-loop\\
$f_1=3$           &         &         &  -1.44  &         &        &cone\\
\hline
\end{tabular}
\end{table}

\vspace{1.5ex}

In table 1 we have used the approximate values of $\eta_f$ to calculate
the cone estimation of the contact exponents and compare these with
corresponding values of a renormalization group calculation.
We use here a perturbation series in terms of an expansion in the
parameter $\varepsilon=4-d$ where $d$ is the space dimension.
In $d=4$ the theory becomes trivial as is intuitively understood, because in
$d=4$ dimensions a random walk (polymer) never meets itself and thus the
nontrivial self-avoiding interaction vanishes. The expansion in $\varepsilon$
is thus an expansion starting from the theory that describes the polymer as a
non-interacting random walk.
The result labeled `1-loop' corresponds to optimal truncation of the series
simply inserting $\varepsilon=1$ or $d=3$ in the first order term of the
expansion. The 3-loop
result includes a resummation procedure that takes into account the asymptotic
nature of the series \cite{Ferhol}. The large $f$ result corresponds to the
cone
approximation.

To summarize, we have shown that the large $f$ approximation for the
short distance force between two star polymers can be consistently fitted
to the results of perturbation theory for low values of the functionality
$f$ of the stars. We have at the same time generalized the approach
to the interaction between two stars of different functionalities $f_1$ and
$f_2$. This is essential in extending the theory of colloidal solutions
of star polymers to general polydispersity in $f$ as it appears naturally
in any real experiment.

\noindent{\bf Acknowledgements}\\
We acknowledge helpful discussions with L.~Sch\"afer.
We thank the Deutsche Forschungsgemeinschaft for support
within SFB 237. Yu.H. gratefully acknowledges support by
Deutsche Akademische Austauschdienst.

\newpage

\end{document}